\documentclass[12pt,a4paper]{article}
\usepackage{amsmath}
\usepackage{amssymb} 
\begin{document}

\begin{titlepage} 
\begin{flushright} IFUP--TH/2023\\
\end{flushright} ~\vskip .8truecm

~  
\vskip 3.0truecm

\begin{center} 
  \Large\bf
Liouville field theory on genus 2 surfaces
\end{center}
\vskip 1.2truecm
\begin{center}
{Pietro Menotti} \\ 
{\small\it Dipartimento di Fisica, Universit{\`a} di Pisa}\\ 
{\small\it 
Largo B. Pontecorvo 3, I-56127, Pisa, Italy}\\
{\small\it e-mail: pietro.menotti@unipi.it}\\ 
\end{center} 
\vskip 0.8truecm
\centerline{May 2023}
\vskip 1.2truecm
                                                              
\begin{abstract}
  Exploiting the generalization of the Weierstrass $\wp$ function to
  genus $2$ given by Komori, we give the exact connection of the
  related monodromy problem for genus $2$ and the classical weak
  $n$-point correlation functions. We also provide the Green function
  of an Helmholtz operator on genus $2$ surfaces and prove the real
  analyticity of the higher genus Green function.  This gives at the
  non perturbative level, through the continuation method, the real
  analyticity of the conformal factor and of the accessory parameters
  in presence of sources.
\end{abstract}

\end{titlepage}
  
\section{Introduction}\label{introduction}

 The classical conformal problem i.e.  the search for a conformal
 factor which provides constant negative curvature to a Riemann
 surface with some prescribed singularities , both of elliptic and
 parabolic type is well explored for genus $0$ topology and also for
 genus $1$
 \cite{picard1,picard2,poincare,lichtenstein,PMaccessory}. Extension
 to supersymmetric formulations have also been given
 \cite{beccaria,FP}. Generally one is interested in bounds,
 perturbative expansions around some given background and analytic
 properties of the solutions when both the position and the strength
 of the singularities varies and for $g>0$ also when the moduli of the
 surface vary.  E.g.  for the sphere topology it has been found that
 the solution are real-analytic function in the position of the
 sources and the result has been extended to genus $1$ including the
 dependence on the moduli \cite{PMelliptic,PMparabolic}. The reason
 for this extension in that we know, along with its detailed
 properties, the Weierstrass $\wp$ function which maps the periodic
 $z$-plane to the two sheeted cut $u$-plane. For $\wp(z)$ we know
 several expressions, the most important being the one in terms of the
 theta function which provides a very quickly convergent
 expansion. For higher genus $g\geq 2$ only qualitative results, like
 existence and uniqueness result of the solutions were known
 \cite{lichtenstein,PMexistence}.  Komori in \cite{komori} gave a
 generalization of Weierstrass's $\wp(z)$ function to genus $2$ expressed as the
 ratio of two $3$-differentials. Each of these differential is
 expressed in terms of a Poincar\'e series. The main point of the
 present paper is to exploit the knowledge of such generalized $\wp$
 function to provide results on the genus $2$ surface which, as is
 well known, is a hyperelliptic surface. Several of the results which
 will be derived in the following sections extend to all hyperelliptic
 surfaces but for $g=2$ we have the advantage of possessing an
 explicit representation of the mapping function.  As is well known
 the non linear problem for the conformal factor can be fruitfully
 translated into a system of linear differential equations in the
 complex plane \cite{deligne,haraoka,cecotti} and the procedure is
 reduced to solving the connection problem. Such a procedure provides
 the value of the accessory parameters in absence of sources. Then one
 can use such a result as a starting point of perturbation theory. We
 shall use it to compute the classical weak $n$-point correlation
 functions on the genus $2$ surface. For the non perturbative
 treatment \cite{poincare} essential is the knowledge of the Green
 functions and of their analytic properties.  Using the explicit form
 of the Green functions we prove, through the continuation method the
 real analyticity of the conformal factor and of the accessory
 parameters.

 The structure of the paper is as follows: In section \ref{generalsec}
 we give a general treatment which holds for all hyperelliptic
 surfaces of the general properties of the mapping function and of the
 connection $Q$ of the theory. In section \ref{komorisec} we give a
 summary of Komori's results and in section \ref{weaknpoint} we use
 the obtained results to develop the perturbation theory in presence
 of weak sources. This implies the computation of the new accessory
 parameters from which on the other hand one obtains the explicit
 expression for the weak semiclassical $n$-point correlation
 function. In section \ref{greenfunctionsec} we compute the Green
 function of an Helmholtz operator on $g=2$ and a give a novel way to
 derive the $g=1$ Green function. We then work out the analytic
 properties of the Green function for $g=2$ and for higher genus. We
 show how such Green functions depend in real analytic way of the
 moduli of the surface. We then give an analytic background field
 $\beta$ which is the starting point for the non perturbative
 treatment of the theory \cite{poincare,PMelliptic,PMparabolic}. The
 result is that for all hyperelliptic surfaces the conformal factor is
 real analytic in the position of the sources and in the moduli of the
 surface and such real analyticity extends to the accessory
 parameters. In section \ref{conclusionsec} we give the conclusions
 and an outlook for further developments.

\section{General treatment: the connection problem}\label{generalsec}

In this section we perform a general discussion of the mapping between
the $z$ representation of the surface and the $(u,w)$ representation
and work out the connection problem i.e. the translation of the problem of
the conformal factor to an ordinary linear differential equation in
the complex plane.

For concreteness we shall refer to the case $g=2$ but
most of the treatment applies also to the general hyperelliptic case.

We start from the representation of the Riemann surface $\cal F$
in the upper half plane $H$. Thus ${\cal F}=H/\Gamma$ is the quotient
of the upper half plane by a properly discontinuous subgroup $\Gamma$
of $SL(2,R)$.

The fundamental domain can be chosen as a curvilinear polygon of $8$
sides ${\rm a}_1~{\rm b}_1~{\rm a}'_1~{\rm b}'_1~{\rm a}_2~{\rm
  b}_2~{\rm a}'_2~{\rm b}'_2$ related by the $SL(2,R)$ transformations
$a_1,b_1,a_2,b_2$ subject to the constraint \break
$a_1b_1a_1^{-1}b_1^{-1 }a_2b_2a_2^{-1}b_2^{-1}=I$.

We shall denote by $z$ the complex coordinate in such half-plane. We
know that the upper half plane is endowed with the  metric
\begin{equation}
  e^{\phi_o} \frac{i}{2}dz\wedge d\bar z =
  \frac{2}{y^2}~dx\wedge dy~,
~~~~~~~~
  e^{\phi_o} =\frac{8}{(z-\bar z)(\bar z-z)}
\end{equation}  
with $z = x+i y$.  It is well known that
there is a fundamental relationship between the conformal factor and
the connection problem for ODE in the complex plane
\cite{deligne,haraoka,cecotti}.  In canonical form
\begin{equation}
  e^{\phi_o} =\frac{8|w_{12}|^2}{(\bar s_2 s_2-\bar s_1 s_1)^2}=
  \frac{8}{(1-\bar f f)^2}\bar f' f'
\end{equation}  
where $w_{12}$ is the Wronskian of the two solutions $s_1,s_2$ of the
equation $s''=0$ and
$f=\frac{s_1}{s_2}$;
\begin{equation}
\Delta \phi_0 = e^{\phi_0}~. 
\end{equation}  

We also have
\begin{equation}
  e^{-\frac{\phi_o}{2}} =\frac{\bar s_2 s_2-\bar s_1 s_1}{\sqrt{8}|w_{12}|}
    =-i\frac{(z-\bar z)}{2\sqrt{2}}
\end{equation}
with 
\begin{equation}
s_1= \frac{1+iz}{\sqrt{2}},~~~~s_2= i\frac{-1+iz}{\sqrt{2}},~~~~w_{12}=1~. 
\end{equation}

We want now to perform the transition from the $z$ representation of
the Riemann surface to the $(u,w)$ representation with
$w^2=4\prod_{j=1}^5(u-e_j)$. 

To this end let $h(z)$ be a meromorphic function on $\cal F$,
i.e. satisfying the correct boundary conditions, with the property
that $h(z)$ is singular only at one point $p$ of $\cal F$ with order
of singularity not higher than the genus i.e. $2$ in our case.  We
know that not for all points $p$ of the Riemann surface there exists a
function with the described properties but this can happen only at a
finite number of points which are the Weierstrass points
\cite{FK}. All surfaces of genus $2$ are hyperelliptic and for a
hyperelliptic surface we have that the number of Weierstrass points is
$2g+2$ \cite{FK}. Thus in our case $g=2$ we have $6$ Weierstrass
points.  We shall denote them by $p_1,~.~.~.~.p_6$.  Then as the
degree of the divisor of $h$ is zero, we have that given $u$ the
equation $u=h(z)$ has two solutions in $z$, counting
multiplicities. The function $h(z)$ provides a holomorphic mapping
between the surface $\cal F$ and $C\cup\infty$ and such a mapping is
of degree $2$.  Thus we know the existence of such function which is
the analogue of the $\wp(z)$ for the torus.

Chosen the point $p_6$ the function $h(z)$ which is regular except for
a pole of order $2$ at $p_6$ is unique up to a linear transformation.
In fact if $h_1(z)$ is an other meromorphic function with a pole of
order $2$ at $p_6$ let us consider
\begin{equation}
h_1(z)- a~h(z)
\end{equation}  
where $a$ has been chosen as to cancel the double pole at $p_6$. Then
as the residue at $p_6$ is zero, such a difference is a meromorphic
function with divisor $1$ and as such it is a constant. Let us
consider now the differential $h'(z)$ whose degree is, according to
the Riemann-Roch theorem, ${\rm deg}(h'(z)) = k(2g-2)=2$ being $k=1$
the order of the differential \cite{FK,jost}.  As the divisor contains
a pole of order $3$ at $p_6$, $h'(z)$ possesses $5$ zeros $q_1
... q_5$ which being the order of the mapping $2$ must be all distinct
and as such they must be simple zeros.  Consider the function
\begin{equation}
h_1(z)=\frac{1}{h(z)-h(q_1)}~.
\end{equation}  
This is a meromorphic function which possess a second order pole in
$q_1$ and no other singularities and thus $q_1$ must coincide with a
Weierstrass point say $q_1=p_1$. Then the equation $u_1= h_1(z)$ becomes
\begin{equation}
u_1= \frac{1}{u-e_1}
\end{equation}  
where $e_1= h(p_1)$ and similarly for the other Weierstrass points.
Thus by changing the Weierstrass point the $u$ variable undergoes
a projective transformation.

We notice how the Weierstrass point $p_6$ of $\cal F$ is mapped to 
$u=\infty$.  We already noticed that the degree of the mapping
$z\rightarrow u$ is of degree $2$. To have a one to one correspondence
between $z$ and the new variables one cuts the $u$-plane from $e_1$ to
$e_2$, from $e_3$ to $e_4$ and form $e_5$ to infinity. Then two
different points of the Riemann surface with the same $u$
belong to different sheets.

Let us consider the relation $z_1\rightarrow z_2$ where $h(z_2)=h(z_1)$.
Near $p_6$ we have
\begin{equation}
h(z)=\frac{1}{(z-p_6)^{2} ~a(z-p_6)}
\end{equation}
with
\begin{equation}
a(z-p_6)=c_0+c_2(z-p_6)^2+ ..
\end{equation}
where we notice that the term $c_1(z-p_6)$ is missing due to the vanishing
of the residue at $p_6$. Then being $a(x)$ a unit we have
\begin{equation}
(z_2-p_6)b(z_2-p_6)=\pm(z_1-p_6)b(z_1-p_6)~.
\end{equation}
With the $+$ sign we have the trivial solution $z_2=z_1$ while we are
interested in the other solution which being $a(x)$ a unit is analytic
in $z_1$. Obviously we have that $z_1\rightarrow z_2$ is an involution.  From
\begin{equation}
h'(z_2)dz_2=h'(z_1)dz_1
\end{equation}
we have that for $h'(z_2)\neq 0$ the dependence of $z_2$ on $z_1$ is
analytic.  If $h'(z_2)=0$ we have that $z_2$ equals a representative
$p''_j$ of some $p_1 . . . p_5$. Then $z_2$ being a double zero of
$h(z_2)-h(z_1)=0$ we have that $z_1$ must be equal to $p_j'$ where
such a point is a representative of $p''_j$.

We notice that
$h'(z_2)$ can vanish only at the representative of $p_1,...p_5$ and there
the equation for $z_2$ near $p_j$ $j=1...5$ becomes
\begin{equation}
(z_2-p''_j)^2a_2(z_2-p''_j)=(z_1-p'_j)^2 a_1(z_1-p'_j)
\end{equation}
whose solutions are
\begin{equation}
(z_2-p_j'') b_2(z_2-p_j'')=\pm(z_1-p_j') b_1(z_1-p_j')
\end{equation}
being still the $b_j$ units and as such they give by the implicit
function theorem the analytic dependence of $z_2$ on $z_1$.  One solution is
the trivial solution $z_2=\gamma z_1$ where $\gamma$ is an element of
the group $\Gamma$ while the other is the one we are interested in.
Then the non trivial solution is given by
\begin{equation}
(z_2-p_j'')b_2(z_2-p_j'')=-(z_1-p_j') b_1(z_1-p_j')
\end{equation}
and thus we have the analytic dependence of $z_2$ on $z_1$ and such a
dependence is extended to the complete $H$ half-plane.  We
notice that both the $a_j(x)$ are of the form $c_0 + c_2
x^2+o(x^3)$ with the absence of $c_1 x$. In fact
consider the meromorphic function $1/(h(z)-e_k)$. Its divisor is
$p_k^{-2},p_6^2$ and thus at $p_k$ we have a second order pole and no
residue which makes in the expansion of $h(z)-e_j$ the third order
term to vanish.

Thus we have the analytic dependence of $z_2$ on $z_1$ also at the
points where $h'$ vanishes. Obviously such dependence is an
involution i.e.  its square is the identity.
Then we have a holomorphic involution which sends the complete $H$
half plane into itself and by a well known theorem it is an element of
$SL(2,R)$.  As $p_6$ is a fixed point such a transformation is an
elliptic transformation of order $2$.

We now move from the coordinates $z$ to $(u,w)$.
We saw how the mapping $z\rightarrow u$ is of degree $2$ with branch
points $p_1 . . . p_6$. Then to have a one to one analytic
correspondence one sets cuts from $e_1$ to $e_2$, from $e_3$ to $e_4$
and from $e_5$ to $\infty$ introducing the variable $w$ with
$w^2=4(u-e_1) . . . (u-e_5)$. The relation between $z$ and $w$ near the
branch point $e_k$ is holomorphic and otherwise points with the same
$u$ are distinguished by the sign of $w$.

Thus $z$ is mapped one to one to the cut plane $(u,w)$ where $w$ is
defined by
\begin{equation}
w^2=4(u-e_1) .~.~.~(u-e_5)~.
\end{equation}  

We have 
\begin{equation}
s_j''=0= s''_j+Q^0_z(z)s_j~~~~{\rm i.e.}~~~~ Q^0_z\equiv 0~.
\end{equation}
We compute now the $Q^0_u$ function which
appears in the differential equation for the basic functions $y_j$.
We know that the functions $y_j$ are forms of order $-\frac{1}{2}$ while
the $Q$ transforms like
\begin{equation}
Q^0_u(u,w) = Q^0_z(z)\big(\frac{dz}{du}\big)^2-\{z,u\}  
\end{equation}
where $\{z,u\}$ is the Schwarz derivative
\begin{equation}\label{schwarzderivative}
\{z,u\} =\big(\frac{dz}{du}\big)^{\frac{1}{2}}~
\frac{d^2}{du^2}\big(\frac{dz}{du}\big)^{-\frac{1}{2}}~.
\end{equation}
Thus the $Q^0$ transforms like a quadratic differential plus the affine term
$-\{z,u\}$.

We shall see below that actually $Q^0_u$ depends only on $u$ and not on $w$.
The most important properties of the Schwarz derivative $\{,\}$ which easily
follow from the group property, are the composition rule
\begin{equation}\label{composotion}
\{g(f(u)),u\}=\{g,f\}(f'(u))^2+\{f,u\}
\end{equation}
from which
\begin{equation}
\{z,u\}du^2 =-\{u,z\}dz^2
\end{equation}
follows.
If $g\in SL(2,C)$ we have 
\begin{equation}\label{sl2c}
\{g(u),u\}= 0
\end{equation}
which follows immediately from eq.(\ref{schwarzderivative}); actually
an $SL(2,C)$ transformation is the only solution of eq.(\ref{sl2c}).

We shall now work out the structure of $Q^0_u$.  In computing
$\{z(u,w),u\}$ we saw that the two values of $z$ corresponding to a
given value of $u$ i.e. $z_1=z(u,w)$ and $z_2=z(u,-w)$ are related by
and elliptic $SL(2,R)$ transformation. Then applying
eq.(\ref{composotion},\ref{sl2c}) we have that
\begin{equation}
Q^0_u(u,w)=Q^0_u(u,-w)
\end{equation}
i.e. $Q^0_u$ does not depend on $w$ or in different words $Q^0_u$ is the
same on both sheets of the cut $u$-plane analytic everywhere except at
the points $u=e_1...e_5$, thus a meromorphic function of $u$.  We
notice that the equality $\{z_2,u\}=\{z_1,u\}$ holds also when $z_1$
and $z_2$ are replaced by any representatives $\gamma_2 z_2$ and $\gamma_1
z_1$ with the $\gamma$'s elements of $\Gamma$ and as such elements of
$SL(2,R)$. The solutions $y_j(u)$ of
\begin{equation}\label{yequation}
y_j''(u)+Q^0_u(u)y_j(u)=0
\end{equation}
are forms of order $-\frac{1}{2}$ i.e. we have
\begin{equation}\label{ysolution}
y_j(u) = s_j(z) \bigg(\frac{dh(z)}{dz}\bigg)^{\frac{1}{2}}~.
\end{equation}

Thus the knowledge of the mapping $u=h(z)$ gives the complete
information on the $Q^0_u(u)$ and the solutions of (\ref{yequation})~.

Exploiting the fact that near $u=e_j$ we have $z-p_j\approx c\sqrt{u-e_j}$
we can compute the coefficient of the singularities which are
\begin{equation}
\frac{3}{16}\frac{1}{(u-e_j)^2} +\frac{b_j}{2(u-e_j)}~.
\end{equation}
Actually from the expansion around $p_j$, $j=1,~.~.~.~5$
\begin{equation}
h(z) = h(p_j)+c_{j2}(z-p_j)^2+ c_{j4}(z-p_j)^4+o((z-p_j)^5 
\end{equation}
we have
\begin{equation}\label{thebj}
b_j= -\frac{3 c_{j4}}{4 c^2_{j2}}
\end{equation}
which are the accessory parameters.  A useful constraint on the
accessory parameters is obtained from the behavior of $h(z)$ for $z$
near $p_6$ i.e. $u\sim(z-p_6)^{-2}$. This correspond to the absence of
charge at the point $u=\infty$. I.e. going over to the uniformizing
variable at infinity $v$ with $v^{-2}=u$ we compute
\begin{equation}
  Q^0_v(v)=Q^0_u(u) \big(\frac{du}{dv}\big)^2-\{u,v\}=
  Q^0_u\big(\frac{1}{v^2}\big) \frac{4}{v^6}-\frac{3}{4 v^2}
\end{equation}
whose regularity at $v=0$ imposes
\begin{equation}\label{uequalinfinity}
Q^0_u(u)\approx\frac{3}{16 u^2}
\end{equation}  
for large $u$. This fact puts two constraints on the five accessory
parameters which are given by
\begin{equation}\label{sumrule1}
b_1+b_2+b_3+b_4+b_5=0
\end{equation}
and
\begin{equation}\label{sumrule2}
  \frac{3}{2}+b_1 e_1+b_2 e_2+b_3 e_3+b_4 e_4+b_5 e_5=0~.
\end{equation}

We conclude that the knowledge of the function $h(z)$
provides the value of the accessory parameters $b_j$; such parameters will
depend on the moduli of the surface. 

Thus for the $Q^0_u$, i.e. in absence of sources, we have
\begin{equation}
Q^0_u(u)=\sum_{j=1}^5 \frac{3}{16}\frac{1}{(u-e_j)^2}+\frac{b_j}{2(u-e_j)}~.
\end{equation}
with the $b_j$ given by (\ref{thebj}).

\section{The Komori construction}\label{komorisec}

In the previous section we gave a general treatment, proving the
existence of the mapping function $h(z)$, unique up to a projective
transformation and of its main properties. The treatment holds for all
hyperelliptic surfaces. On the other hand one should like to have an
explicit representation of the such a function. For $g=2$ such a
representation has been provided by Komori \cite{komori} as the ratio
of two 3-differentials each of them given by a Poincar\'e series.
Such a function is the natural extension to genus $2$ of the
Weierstrass $\wp(z)$ function for the torus.
In this section we summarize the construction by Komori.

One starts from the presentation of the genus $2$ fuchsian group
\cite{FK,jost}
\begin{equation}
  \Gamma:=\langle a_1,a_2,b_1,b_2|a_1 b_1 a_1^{-1} b_1^{-1}a_2 b_2 a_2^{-1}b_2^{-1}=
  I\rangle\subset
      {\rm PSL}(2,R)
\end{equation}
and considers the intersections $z_i$ of the hyperbolic
axis of $a_i$, $b_i$ and one denotes with $\delta_1, \delta_2$ the
elliptic elements of order two of ${\rm PSL}(2,R)$ whose fixed points
are $z_1,z_2$. One proves \cite{komori} that the group $G$ generated
by $\Gamma$ and $\delta_1$ is a Fuchsian group of genus zero
containing $\Gamma$ as subgroup of index $2$ i.e.
\begin{equation}
G=\Gamma\cup \Gamma\delta_1
\end{equation}
and that the group $G$ possesses the presentation
with six generators
\begin{equation}\label{Gpresentation}
G=:\langle
\gamma_1\dots\gamma_6|\gamma_1^2=\dots\gamma_6^2=
\gamma_1\gamma_2\gamma_3\gamma_4\gamma_5\gamma_6=I
\rangle
\end{equation}
where the $\gamma_k$'s are given by the six $a_i\delta_i$, $\delta_i$,
$b_i\delta_i$, $i=1,2$. From the presentation (\ref{Gpresentation}) we
have that the genus of ${\cal S}\equiv H/G$ is zero
\cite{beardon,long,shvartsman}.

Komori considers a $G$-automorphic meromorphic 3-differential on
$\cal S$ i.e. a meromorphic $\hat g$ such that under the
transformation $v'=\gamma v$
\begin{equation}
\hat g(\gamma v) \gamma'(v)^3 = \hat g(v) 
\end{equation}
for all $\gamma\in G$. Such form has degree ${\rm deg}(\hat g) = k(2
g-2)$ where $g$ is the genus; in our case as $g=0$ and $k=3$ we have
${\rm deg}(\hat g)=-6$.  Given the holomorphic mapping of $\cal F$ to
$\cal S$ we can pull back the 3-differential $\hat g$ on $\cal S$ to
the 3-differential $g$ on $\cal F$ according to
\begin{equation}\label{hatgtog}
g(z) = \hat g(v) \big(\frac{dv}{dz}\big)^3~.
\end{equation}
With regard to the $3$-differential $\hat g$ on $\cal S$ we have that
in the neighborhoods of the fixed points $p_k$ of $\gamma_k\in G$ one
cannot use the coordinate $z$ but some coordinate $v$ which being the
mapping $S\rightarrow {\cal F}$ of order $2$ is related using
Weierstrass preparation theorem by
\begin{equation}
(z-p_k)^2 + a_1(v)(z-p_k)+ a_0(v)=0~.
\end{equation}
The degree of $g(z)$ is now $3(4-2)=6$ as the genus of $\cal F$ is $2$.
If we
demand that $\hat g(v)$ be such that the ensuing $g(z)$, see 
eq.(\ref{hatgtog}), be holomorphic and not simply meromorphic, we have
that the poles of $\hat g$ must lye at the fixed points $p_k$ and
thus the zeros of $g(z)$ are all distinct and thus of first order.

Let us consider next a meromorphic $G$-automorphic $3$-form $f(z)$
whose poles are just a simple pole at the orbit of the point $p_6$.
Using ${\rm deg}(f)= 6$ it is not difficult to prove that there are only
two possible configurations for its divisor
\begin{equation}
(f) = p_1~p_2~p_3~p_4~p_5~p_6^{-1}~q_1~q_2 ~~~~(I),~~~~~~~~
(f) = p_{j_1}~p_{j_2}~p_{j_3}~p_{j_4}~p^3_{j_5}~p_6^{-1}~~~~(II)~.
\end{equation}
The ratio $h(z)=f(z)/g(z)$ is a function on $\cal F$.  In
configuration $(I)$ $h(z)$ it has a double pole at $p_6$ and two
simple zeros at $q_1,q_2$, while in configuration $(II)$ $h(z)$ has a
double pole at $p_6$ and a double zero at $p_{j_5}$ in both cases
satisfying the requirement of the function $h(z)$ of the previous
section.  $g(z)$ can be written as the Poincar\'e series
\cite{komori}
\begin{equation}\label{gpoincseries}
g(z)=\sum_{\gamma\in G}P(\gamma z)\gamma'(z)^3
\end{equation}
where $P$ is a properly chosen function holomorphic in the upper half
plane $H$.  If the Poincar\'e series (\ref{gpoincseries}) converges to
a non identically zero function it follows due to the presence of
$\gamma'(z)^3$ that $g(z)$ is a holomorphic $G$-automorphic
differential of order $3$.  For $f(z)$ then it follows immediately
that the series
\begin{equation}\label{fpoincseries}
f(z)=\sum_{\gamma\in G}\frac{1}{\gamma z -p_6} P(\gamma z)\gamma'(z)^3
\end{equation}
also convergences and defines a $G$-automorphic meromorphic
$3$-differential with a simple pole at $z=p_6$. Then the ratio
\begin{equation}\label{hratio}
  h(z)=\frac{f(z)}{g(z)}
\end{equation}
will be a meromorphic function on $\cal F$ with only one second order
pole at $p_6$ and two zeros which may occasionally coincide.
The main point in this constructions is the choice of a $P$
holomorphic in $H$ and such that the Poincar\'e series
(\ref{gpoincseries}) converges to a not identically zero function.

The function $P(z)$, analytic in the upper half plane $H$ is given by
\cite{komori}
\begin{equation}
P(z) = \frac{1}{z-\gamma_1\nu_1}\prod_{j=1}^5\frac{1}{z-\nu_j}
\end{equation}
where $\nu_j$ is the attractive end point of the axis of the
hyperbolic element $\gamma_j\gamma_6$, with the convention that if
$\gamma_1\nu_1=\infty$ or $\nu_j=\infty$ the terms $z-\gamma_1\nu_1$
or $z-\nu_j$ are dropped from the formula.

Thus eq.(\ref{hratio}) provides an explicit form of the function $h$
with which we were dealing in section \ref{generalsec} i.e. a
function on $\cal F$ with a single pole of order not higher than $2$
and no other singularity.

\section{The weak n-point function}\label{weaknpoint}

The advantage of the representation $(u,w)$ of the Riemann surface is
that we can insert, by modifying the $Q^0_u(u)$, sources satisfying
the correct boundary conditions, something that would be very difficult
to obtain in the $z$ representation.

We shall start by introducing a single weak source at the point
$(t,w_t)$ of the surface.
\begin{equation}\label{fullQ}
  Q(u,w) = Q^0_u(u)+ \epsilon ~q(u,w)
\end{equation}
where see \cite{PMtorusII,PMpoly} 
\begin{equation}\label{smallq}
  q(u,w)=
  \frac{(w+w_t)^2}{4(u-t)^2 w^2}+\frac{b_t(w+w_t)}{4(u-t)w}
  +\sum_{j=1}^5 \frac{b_j'}{2(u-e_j)}~.
\end{equation}
We note that the additional term $q$ in (\ref{fullQ}) is a true quadratic
differential also at the non perturbative level i.e. we have
\begin{equation}\label{qdifferentail}
q(u,w)\bigg(\frac{dh}{dz}\bigg)^2= q_z(z)~. 
\end{equation}
The factors $w+w_t$ in (\ref{smallq}) project the singularities on
the correct sheet.  We notice that in order to satisfy the
monodromy conditions the original accessory parameters $b_j$ are
modified by the addition of the terms ~$\epsilon ~b_j'$ which we shall
compute. Also the constraints (\ref{sumrule1},\ref{sumrule2}) extend
to the case when sources are present and thus in addition to
(\ref{sumrule1},\ref{sumrule2}) we have now
\begin{equation}\label{newsrule1}
b'_1+b'_2+b'_3+b'_4+b'_5+\frac{b_t}{2}=0
\end{equation}
and
\begin{equation}\label{newsrule2}
\frac{1}{2}+b'_1 e_1+b'_2 e_2+b'_3 e_3+b'_4 e_4+b'_5 e_5+\frac{b_t}{2} t=0~.
\end{equation}
Then as expected one has the same number of constraints and to
determine the new accessory parameters we needs $6-2=4$ variations of
accessory parameter to be obtained by imposing the $SU(1,1)$ nature
of the monodromies along non contractible cycles.

To first order the perturbed solution $y^c_j=y_j+\epsilon \delta y_j$
is given by

\begin{equation}
  Y^c =\begin{pmatrix}
  y^c_1\\
  y^c_2
  \end{pmatrix}=
  \begin{pmatrix}
  y_1+\epsilon\delta y_1 \\
  y_2+\epsilon\delta y_2
  \end{pmatrix}
  =(1+\epsilon F)Y=
  (1+\epsilon F)
  \begin{pmatrix}
  y_1\\
  y_2
  \end{pmatrix}
\end{equation}  
where
\begin{equation}\label{Fmatrix}
  F=
  \begin{pmatrix}
    \frac{I_{12}}{w_{12}}+c_{11} &-\frac{I_{11}}{w_{12}}+c_{12}\\
    \frac{I_{22}}{w_{12}}+c_{21} &-\frac{I_{12}}{w_{12}}+c_{22}
  \end{pmatrix}
\end{equation}
and
\begin{equation}\label{Iintegrals}
I_{jk}(u) = \int_{u_0}^u q(u,w) y_j(u) y_k(u) du~.
\end{equation}  
It is useful to keep the value of the Wronskian unchanged which is
obtained with $c_{11}+c_{22}=0$.  We shall denote with $\tilde Y$ and
$\tilde F$ the value of the given quantities $Y, F$ computed along a
path which encircles a given singularity or cut.

For the unperturbed problem, which is already solved by the
$y_j(u)$ see eq.(\ref{ysolution}), we have
\begin{equation}
\tilde Y = M Y,~~~~~~M\in SU(1,1)~.
\end{equation}
$M$ does not depend on the point of the comparison but only on the circuit
which embraces a chosen singularity.

For the perturbed one, to first order we have
\begin{equation}
  \tilde Y^c = (1+\epsilon\tilde F)MY=(1+\epsilon\tilde F)M(1-\epsilon F)Y^c=
  M^c Y^c = (M+\epsilon\delta M)Y^c
\end{equation}
where
\begin{equation}\label{deltaM}
\delta M= \tilde F M- M F~.
\end{equation}  

It is easily proven that
\begin{equation}
\tilde F'=MF'M^{-1} 
\end{equation}  
and thus we have
\begin{equation}
\delta M' = \tilde F'M-M F'=0
\end{equation}  
i.e. also $\delta M$ does not depend on the point.

As already explained we shall impose that $M^c=M+\epsilon\delta M\in
SU(1,1)$ by fixing the parameters $b'_1\cdot\cdot\cdot b'_5$ and
$b_t$.  The request applies to the closed contours $a_1,a_2,b_1,b_2$
and to a contour around the point $(t,w_t)$. With regard to $a_i$ and
$b_i$ their images on the $(u,w)$ plane can be chosen as follows:
$a_1$ lies on the first sheet and embraces the cut $(e_1,e_2)$ while
$a_2$ lies again on the first sheet and embraces the cut
$(e_3,e_4)$. $b_1$ embraces the branch points $e_2,e_5$ lying for the
first half on the first sheet and for the second half of the second
sheet while $b_2$ embraces the points $e_4,e_5$ again lying for the
first half on the first sheet and for the second half on the second
sheet \cite{FK}.

\bigskip

For the monodromy around the point $(t,w_t)$ we have
\begin{equation}
  M(t) =
  \begin{pmatrix}
    1+\epsilon\frac{\delta I_{12}(t)}{w_{12}} & -\epsilon\frac{\delta
      I_{11}(t)}{w_{12}} \\
    \epsilon\frac{\delta I_{22}(t)}{w_{12}} &
    1-\epsilon\frac{\delta I_{12}(t)}{w_{12}}
  \end{pmatrix}~.
\end{equation}  
The $\delta I_{jk}(t)$ are the integrals $I_{jk}$, eq.(\ref{Iintegrals}).
computed along a closed path surrounding the point $(t,w_t)$.
\begin{equation}
  \delta I_{jk}(t) =
  i \pi (b_t -T) y_j(t)y_k(t)+2\pi i (y_j(t)y_k(t))'
\end{equation}
where
\begin{equation}
T = \sum_{j=1}^5\frac{1}{t-e_j}~.
\end{equation}
Impose now $M_{12}=\overline {M}_{21}$ i.e.  
\begin{equation}
  (b_t-T)y^2_1(t)+2(y^2_1(u))'_t=
  (\bar b_t-\bar T)\bar y^2_2(t)+2\overline{(y^2_2(u))'_t}
\end{equation}
and  $M_{11} = \overline{M}_{22}$ to reach
\begin{equation}
  (b_t-T)y_1(t)y_2(t)+2(y_1(u)y_2(u))'_t=
  (\bar b_t-\bar T)\overline{y_1(t) y_2(t)}+2\overline{(y_1(u)y_2(u))'_t}~.
\end{equation}
Multiplying the first by $\bar y_1$ and the second by $\bar y_2$ and
subtracting we obtain
\begin{equation}\label{btT}
  b_t-T = -4\frac{y_2'\bar y_2 -y_1'\bar y_1 }{ y_2\bar y_2 -
    y_1\bar y_1 }\bigg|_t~.
\end{equation}
The $y_j$ and their derivatives  are known explicitly.
Thus the accessory parameter $b_t$ can be determined independently of
the variation $b'_j$ of the other accessory parameters.

\bigskip

We must now prove how the imposition of the monodromies along the
loops $a_1,b_1,a_2,b_2$ fixes the $b'_j$ and the $c_{jk}$. As the
$b_t$ is already known exploiting the relations
(\ref{sumrule1},\ref{sumrule2}) we need to fix only three of the five
accessory parameters $b'_j$. With regard to the $c_{jk}$ we notice
that the solution of the monodromy problem contains a free global
$SU(1,1)$ transformation and thus the matrix $c_{jk}$ is determined up
to an element of $su(1,1)$. Through such
a transformation we can always bring the $c_{jk}$ to the form
$c_{11}=-c_{22}=r_1={\rm real}$ and $c_{12}=r_2+i r_3$ $c_{21}=-r_2+i
r_3$ with real
$r_j$. We have thus three real degrees of freedom in the matrix
$c$. The imposition of the $SU(1,1)$ nature of the new
transformation requires to $\delta M_{12}= \overline{\delta M_{21}}$
which are two real constraints; if we require the further
real constraint $|\delta M_{11}| = |\delta M_{22}|$ we have, due to
$M+\epsilon\delta M\in SL(2,C)$, the consequent $SU(1,1)$ nature of the
transformation $M+\delta M$. I.e. three real constraints assures the
monodromic nature of the perturbed transformation.

We start now from the three cycles
$a_1,~b_1,~ a_2$ with $3\times 3=9$ conditions and $3\times 2+ 3=9$
free parameters i.e. the real and imaginary parts of the $b'_j$ and
the $c_{jk}$ parameterized as described above.

The change in the monodromy matrix $M$ around a circuit $a_1$ which
embraces the cut $(e_1,e_2)$ is given by eq.(\ref{deltaM}) with $F$
provided by eq.(\ref{Fmatrix}). Thus $F$ contains the integral from
$u_0$ to the starting point of the loop and $\tilde F$ also the
integral along the loop reaching at the end the same $(u,w)$
point. Such integral is provided by
\begin{equation}
\oint_{a_1} q(u,w) y_j(u) y_k(u) du = \int^{z_2}_{z_1} q_z(z) s_j(z)
s_k(z) dz
\end{equation}  
where $q_z(z)$ is the quadratic differential (\ref{qdifferentail})~.
The starting point and the endpoint in the second integral are
equivalent points on the border of the $\cal F$ region one lying of the
side $b_1$ and the other on the side $b_1^{-1}$ or more simply along
the loop $a_1$. Notice that such integral is well defined as $h(z)$ is
a single valued function.  The $\delta M$ relative to the four
circuits $a_1,a_2,b_1,b_2$ are linear expressions in
$b_1',b_2',b'_3$ and in the $c_{jk}$.

With regard to the last cycle $b_2$ we have,
starting from the loop of the monodromies which goes along
the sides $a_1,b_1,a_1^{-1},b_1^{-1},a_2, b_2,a_2^{-1},b_2^{-1}$ and
contracting it to a circle around the source at $z_t$
\begin{equation}\label{Mloop}
M(q_t) M(a_1) M(b_1) M(a^{-1}_1)M(b^{-1}_1)M(a_2) = M(b_2) M(a_2) M(b^{-1}_2)
\end{equation}
where all the $M$ are elements of $SL(2,C)$. Eq.(\ref{Mloop}) with
$M(q_t), M(a_1),M(a_2),M(b_1)\in SU(1,1)$ imposes the change in
$M(b_2)$ i.e. $M(b_2)\rightarrow M(b_2)(1+\epsilon f)$ with $f\in su(1,1)$
thus realizing the monodromy condition at all cycles.

One should not confuse the $a_j,b_j\in SL(2,R)$ with the
$M(a_j),~M(b_j)\in SL(2,C)$ associated with them which through special choice of
the accessory parameters become elements of $SU(1,1)$ thus
assuring the single valuedness of the conformal factor.

\bigskip

The action in the $u$-representation $S_u$ taking into account
the behavior (\ref{uequalinfinity}) is given by \cite{PMpoly,PMtorusII}
\begin{eqnarray}\label{SuactionTorus}
S_u&=& 
\frac{1}{2\pi}\int_{D_\varepsilon}(\frac{1}{2}\partial\varphi\wedge\bar
\partial\varphi+ e^\varphi du\wedge d\bar u)
\frac{i}{2}\nonumber\\
&-&\frac{\eta_t}{4\pi i}\oint_{\varepsilon_t} \varphi(\frac{du}{u-u_t} -\frac{d\bar
  u}{\bar u-\bar u_t})-\eta_t^2\log\varepsilon_t^2\nonumber\\
&-&\frac{1}{16\pi i}\oint^d_{\varepsilon_l}\varphi(\frac{du}{u-e_l} 
-\frac{d\bar  u}{\bar u-\bar e_l}) 
-\frac{1}{8}\log\varepsilon_l^2\nonumber\\
&+&\frac{1}{8\pi i}\frac{3}{2}\oint^d_{R_u}\varphi(\frac{du}{u} 
-\frac{d\bar  u}{\bar u}) 
+\frac{1}{2}\big(\frac{3}{2}\big)^2\log R_u^2
\end{eqnarray}
where $D_\varepsilon$ is the double sheeted plane 
and the index $d$ on the contour integrals means that a double turn 
has to be taken around the kinematic
singularities $e_l$, $l=1~.~.~.~5$ and at $\infty$ in order to come back
to the starting point.

We come back now to the relation (\ref{btT}). It can be written as
\begin{equation}
  b_t-T = -4\frac{y_1'\bar y_1 -y_2'\bar y_2 }{ y_1\bar y_1 -
    y_2\bar y_2 }\bigg|_t = 2 \frac{\partial \varphi(u)}{\partial u}\bigg|_{t}~. 
\end{equation}  
We recall now the general non perturbative relation which follows from 
eq.(\ref{SuactionTorus})
\begin{equation}\label{Xequation}
\frac{\partial S_u}{\partial \eta_t}= -X_t
\end{equation}  
where $X_t$ is the finite part of the conformal factor at $t$ i.e.
\begin{equation}
\varphi (u) = -2\eta_t\log|u-t|^2 + X_t +o(u-t)~.
\end{equation}
Notice that in the limit $\eta_t=\epsilon\rightarrow 0$ we have
$X_t = \varphi(t)$.
On the other hand Polyakov relation tells us
\begin{equation}
  \frac{\partial S_u}{\partial t} = -\frac{B_t}{2} = -\epsilon
  \frac{(b_t -T)}{2} = -\epsilon \frac{\partial \varphi^0} {\partial t}~.
\end{equation}
For the proof of Polyakov relation for genus $g>0$ surfaces see
\cite{PMtorusblocks} Appendix A.

Then integrating in $t$ we have
\begin{equation}
  S_u = S_u^0 -\epsilon \varphi^0(t) + \epsilon f+O(\epsilon^2)
\end{equation}  
where $f$ can depend on the moduli $e_j$ and on the other sources.
Then we apply eq.(\ref{Xequation})
\begin{equation}
  \frac{\partial S_u}{\partial \epsilon}
  =-\varphi^0(t)+f = -\varphi^0(t)
\end{equation}  
from which $f=0$ thus reaching the result $S_u=S_u^0-\epsilon
\varphi(t)$.  Exploiting now the additivity of perturbation theory to
first order we can extend the result to $n$ weak sources
reaching
\begin{equation}\label{weakcorr1}
\langle V_{\epsilon_1}(t_1)\dots V_{\epsilon_n}(t_n)\rangle
 = e^{-S^0} e^{\sum_{j=1}^n\epsilon_j\varphi^0(t_j)}~.
\end{equation}
Equation (\ref{weakcorr1}) was conjectured for the sphere topology in
\cite{ZZ} and proven for the sphere topology in \cite{MV}.
Here it is proven for all hyperelliptic surfaces. In addition in the case
$g=2$ thanks to the knowledge of the function $h(z)$ we know it explicitly
as
\begin{equation}\label{weakcorr2}
\varphi^0(u,w)=-\log|z-\bar z|^2+3 \log 2 -\log\bigg|\frac{dh}{dz}\bigg|^2~.
\end{equation}
Coming back to the single source insertion we notice that the real
analytic dependence of the solution extends from the dependence on the
source position to the strength of the singularities \cite{PMelliptic}.
The result of \cite{PMaccessory} and \cite{PMelliptic} assures us that
the perturbative series in $\epsilon$ converges
for $|\epsilon|<\frac{1}{4}$ being $\frac{1}{4}$  the parabolic
limit.

\section{The Green functions and the real analyticity of the conformal factor}
\label{greenfunctionsec}

In the previous section we determined the conformal factor of genus
$2$ surfaces when $n$ weak sources are present. Thus while in absence of
sources we have
\begin{equation}\label{nosource}
4\partial_z\partial_{\bar z}\phi= e^\phi
\end{equation}
or
\begin{equation}
  \Delta_{LB}\phi=1~,
\end{equation}  
in presence of a source we have
\begin{equation}
4\partial_z\partial_{\bar z}\phi_\epsilon= e^{\phi_\epsilon}+ \epsilon\delta(z-z_t)~.
\end{equation}
Taking the derivative w.r.t. $\epsilon$
\begin{equation}
4\partial_z\partial_{\bar z}G(z,z_t)= e^{\phi}G(z,z_t)+\delta(z-z_t)~.
\end{equation}
i.e. we found the Green function of the Helmholtz operator
$\Delta_{LB} -1$.

From the previous section we have that
\begin{equation}
\varphi_\epsilon=-\log \frac{(Y^c,\eta Y^c)^2}{8 |w_{12}|^2}
\end{equation}
where $\eta={\rm diag}(-1,1)$ and thus
\begin{equation}
  \frac{\partial \varphi_\epsilon}{\partial \epsilon}=
  -2\frac{(Y,(F^+\eta+\eta F)Y)}{(Y,\eta Y)}
\end{equation}
and $F^+\eta+\eta F=m$ is the hermitean matrix with
\begin{eqnarray}
&& m_{11}=-2{\rm Re}\bigg(\frac{I_{12}}{w_{12}}+ c_{11}\bigg)= -2{\rm
    Re}\bigg(\frac{I_{12}}{w_{12}}+ r_1\bigg)=m_{22}\\ &&
  m_{12}=\frac{I_{11}}{w_{12}}+ \frac{\bar I_{22}}{w_{12}}-c_{12}+\bar
  c_{21}=\frac{I_{11}}{w_{12}}+\frac{\bar I_{22}}{w_{12}}-2(r_2+ir_3)=\bar m_{21}
\end{eqnarray}
i.e.
\begin{equation}
G(z,z_t) =-\frac{e^{\frac{\varphi}{2}}}{\sqrt{2}|w_{12}|}  (Y,mY)~.
\end{equation}
A similar Green function for the sphere topology in presence of
background charges was derived in \cite{MV}.

\bigskip

On a compact Riemann surface the Green function is defined
\cite{royden} as a function $g$ of the point $p$ which is harmonic in
${\cal F}\backslash(q,q_0)$ and near $q$ and $q_0$ is given by
\begin{equation}
  \frac{1}{2\pi}\log|z-z_q| + c(z), ~~~~~
  -\frac{1}{2\pi}\log|z-z_{q_0}| + c_0(z)
\end{equation}  
with $c,c_0$ $C^\infty$ functions and $z=z_p$ is a local uniformizing
variable. In addition it is requested that $g(p,p_0;q,q_0)$ vanishes
at some $p=p_0\neq q,q_0$. It is not difficult to prove \cite{royden}
that such function exists, is unique, is real and that we have
\begin{equation}
g(p_0,p;q,q_0)=-g(p,p_0;q,q_0),~~~~~~~~g(q,q_0;p,p_0)=g(p,p_0;q,q_0)~.
\end{equation}  
It satisfies 
\begin{equation}
  4\partial_z \partial_{\bar z} g(p,p_0;q,q_0)=
  \delta(z-z_q)-\delta(z-z_{q_0})~.
\end{equation}  
We can multiply to the left by the conformal factor $e^{-\phi(z)}$
to obtain
\begin{equation}\label{twosources}
\Delta_{LB} ~g(p,p_0;q,q_0)=
  \delta^\phi(p,q)-\delta^\phi(p,q_0)
\end{equation}
where $\Delta_{LB}$ is the Laplace-Beltrami operator. We can integrate the
previous equation (\ref{twosources}) in
$e^{\phi(q_0)} dq_0$ over the whole compact surface to obtain
\begin{equation}\label{standardGreen}
\Delta_{LB}~   g(p,p_0;q)= \delta^\phi(p,q)-\frac{1}{V}
\end{equation}
with
\begin{equation}
  g(p,p_0;q)= \frac{1}{V}\int g(p,p_0;q,q_0) e^{\phi(q_0)} dq_0
\end{equation}
where $V$ is the total volume of the surface and $ g(p,p_0;q)$
vanishes at $p_0$. As is well known the appearance of $1/V$ on the
r.h.s. of eq.(\ref{standardGreen}) is due to the vanishing of the
total charge on a compact surface.

\bigskip

The knowledge of the Green function and of its analytic properties
is fundamental for the non perturbative treatment of the Liouville
equation \cite{poincare}. In fact starting from the Green function the
problem is solved at the non perturbative level \cite{poincare}
in terms of uniformly convergent series for any number of sources,
obviously satisfying the Picard bounds. In
\cite{PMelliptic,PMparabolic} it has been shown how the real-analytic
properties of the Green function is inherited by the full solution.

The Green function on the Riemann sphere is well known.
In the case of the torus $g=1$ the equation (\ref{nosource}) has no
solution in absence of sources. Despite that one can apply the
perturbation theory developed in the previous section using
the conformal factor
\begin{equation}
e^\varphi= \frac{8|w_{12}|^2}{(\bar y_2 y_2 - k^2 \bar y_1 y_1)^2}
\end{equation}
with
\begin{eqnarray}
  && y_2 = \Pi^{\frac{1}{4}}+\epsilon \delta y_2\\
  && y_1 = \Pi^{\frac{1}{4}} Z(u)\\
  && k^2 = c ~\epsilon 
\end{eqnarray}
where
\begin{equation}
  \Pi(u) = 4(u-e_1)(u-e_3)(u-e_3),~~~~~~~~
  Z(u) = \int_{e_3}^u\frac{du}{\Pi^{\frac{1}{2}}(u)}~.
\end{equation}  
The result for $e^{\varphi}$ can be obtained by a single quadrature
\cite{PMtorusI} in
term of the $\theta_1$ function and its derivative w.r.t. $\epsilon$
gives the well known solution of
\begin{equation}
4\partial_z\partial_{\bar z} G(z) = \delta(z)  - \frac{1}{\tau^I}
\end{equation}
i.e. 
\begin{equation}\label{torusgreen}
G(z) =\frac{1}{4\pi}\log\big|\theta_1(z|\tau)|^2+\frac{(z-\bar
  z)^2}{8\tau^I} ~.
\end{equation}

\bigskip

On a compact Riemann surface of genus $g\geq 1$, which includes the
previous case, we have \cite{DPh,divecchia1,divecchia2}
\begin{equation}\label{GB}
  G_B(z,w) = \frac{1}{8\pi}\log\bigg(|E(z,w)|^2 e^{-2\pi{\rm
    Im}(\int_{w}^{z}\omega_j)~({\rm
    Im}\Omega)^{-1}_{jk}~{\rm Im}(\int_{w}^{z}\omega_k)}\bigg)
\end{equation}
satisfying
\begin{equation}
4\partial_z\partial_{\bar z}G_B(z,w)=\delta(z-w)- \frac{1}{2\pi} K_g(z)
\end{equation}
and
\begin{equation}
  K_g(z)=\pi\sum_{j,k=1}^g\omega_j(z)({\rm Im}\Omega)^{-1}_{jk}~
    \overline{\omega_k(z)}~.
\end{equation}
$E$ is the prime form
\begin{equation}
  E(z,w)= \frac{\theta[\delta][\int_w^z \omega,\Omega]}
  {\sqrt{\omega_\delta(z)}\sqrt{\omega_\delta(w)}}
\end{equation}
where
\begin{equation}
  [\delta]=
  \begin{bmatrix}\delta'\\ \delta''
   \end{bmatrix}~~~~~~~~{\rm with}~~~~ 4\delta'\delta''={\rm odd}
\end{equation}
and the $\theta[\delta]$ is the Riemann theta function with
characteristic $\delta$.  The abelian differentials have been
normalized as \cite{DPh,FK,jost}
\begin{equation}\label{omeganormalization}
  \int_{a_k}\omega_j = \delta_{kj},~~~~~~~~\int_{b_k}\omega_j = \Omega_{kj}
  =\Omega_{jk}
\end{equation}
and $\omega_\delta(z)$ is the abelian differential
\begin{equation}
  \omega_\delta(z) = \sum_j \frac{\partial\theta[\delta][0,\Omega]}{\partial
  \zeta_j}\omega_j(z)~.
\end{equation}

Under cycles $a_k$, $E$ is left invariant up to a sign, while under cycles
$b_k$ $E(z,w)$ goes over to \cite{DPh}
\begin{equation}
E(z,w)\rightarrow -E(z,w) e^{(-i\pi\Omega_{kk}-2\pi i\int_w^z\omega_k)}~.
\end{equation}
The exponential part in the expression (\ref{GB}) of $G_B$ is there to cancel
such a multiplicative dependence.

The function $K_g$ does not play any role in the procedures of
\cite{poincare,PMelliptic,PMparabolic} as there one works always under
the condition of zero total charge for the sources.

We come now to the real analytic dependence of the conformal factor
and of the accessory parameters on the source positions and on the
moduli of the surface.  Real analyticity of the accessory parameters
on the position of the singularities has been proven by Kra
\cite{kra} for the sphere topology, in presence of only parabolic
and finite order elliptic singularities. The result was extended in
\cite{PMelliptic,PMparabolic} in presence of general elliptic and
parabolic singularities for the sphere topology and for the torus.
Such real-analytic dependence of the conformal factor and of
the accessory parameters plays a pivotal role in the developments of
conformal theories. We recall that the conformal factor can be
obtained at the non perturbative level in the form of uniformly
convergent series \cite{poincare} starting from a background field
$\beta(z)$ and the Green function of the Laplace operator.

The proof of the real analytic nature of the conformal
factor in the position of the sources and on the moduli rests of the
real analyticity of the background $\beta$ and of the Green function. 
The starting point is the construction of a background $\beta$ which
depends real-analytically on the position of the sources and for
$g\geq 1$ depends real-analytically on the moduli.

The background field $\beta$ must satisfy near the singularities the
requirements
\begin{equation}
0<\lambda_m<\beta |z-z_k|^{-4\eta_k}<\lambda_M
\end{equation}
and be elsewhere bounded \cite{lichtenstein,PMexistence}.

For $g=1$ with elliptic singularities one can use for $\beta$
\begin{equation}
\beta =c\prod_k [\wp(z-z_k)\overline{\wp(z-z_k)}+p^2]^{\eta_k}
\end{equation}
which near each singularity behaves as
\begin{equation}
[(z-z_k)(\bar z-\bar z_k)]^{-2\eta_k}~.
\end{equation}
Moreover the $\beta$ is subject to the constraint \cite{PMexistence}
\begin{equation}
  \int\beta(z)d^2z=4\pi\big(\sum_k2\eta_k+2(g-1)\big)
\end{equation}
which is always possible due to the topological constraint
\begin{equation}
\sum_k 2\eta_k > 2(1-g)~.
\end{equation}
In this way $c$ becomes a real analytic function of the position of
the sources \cite{PMelliptic}. For $g=2$ and higher we can use the
method developed in \cite{PMparabolic} writing
\begin{equation}\label{rhoreg}
\beta = c \prod_k\bigg(\rho(|z-z_k|)~|z-z_k|^{-2}+1\bigg)^{2\eta_k}
\end{equation}
where $\rho$ is a positive $C^\infty$ function, which equals $1$ in the
neighborhoods of $0$ and such that for different $k$ the support of
the $\rho$'s in (\ref{rhoreg}) do not overlap.  Then one can work as
explained in \cite{PMparabolic}. One looses in
the proof the analyticity of
the conformal factor near the sources but then one can exploit the
uniqueness theorem \cite{lichtenstein,PMexistence} and the freedom in
the $\rho$ to recover analyticity everywhere except at the
sources. Similarly one can work in presence of parabolic singularities
\cite{PMparabolic}.

With regard to the real-analytic dependence of the Green function on
the moduli of the surface, for the torus it is explicit from the expression
(\ref{torusgreen}). For genus $2$ and higher in the process of
constructing the Green function (\ref{GB}) we must first perform the
transition to the normalized abelian differentials $\omega_j$
(\ref{omeganormalization})  starting from
\begin{equation}
\frac{du}{w},~~\frac{u~du}{w},~\dots \frac{u^{g-1}~du}{w}~.
\end{equation}
This transition preserves the analyticity in the moduli of the
surface. Similarly $E$ depends analytically on the moduli and thus
the Green function (\ref{GB}) is real analytic in the moduli.

Thus following the procedures of \cite{PMelliptic,PMparabolic} we
have that the conformal factor depends in real analytic way
on the source positions and on the moduli of the surface for all hyperelliptic
surfaces. The same conclusion applies to the accessory parameters.

\section{Conclusions}\label{conclusionsec}

In this paper we have exploited the generalization of the Weierstrass
$\wp$ function to genus $2$ given by Komori \cite{komori} to derive
results on genus $2$ surfaces.  The knowledge of such generalization,
in addition to the exact connection, provides the starting point to
the development of perturbation theory and thus to the computation of
correlation functions. As an application we gave the explicit
expression of the $n$-point weak correlation function. We provide also
the Green function of an Helmholtz operator on genus two surface and
work out the analytic properties of the Green functions for $g\geq
2$. This knowledge is used to prove the real-analytic nature of the
dependence of the conformal factor on the source positions and on the
moduli for higher genus surfaces. The real analyticity of the
accessory parameters easily follows.

For future developments one may try the extend Komori's procedure to
give an explicit form of the mapping function for $g=3$ and possibly
higher $g$ hyperelliptic surfaces.  For non hyperelliptic surfaces an
inroad to the case of genus $3$ has been given in \cite{eilbeck}.

\eject

\vfill


\end{document}